\documentclass{chi2009}
\usepackage{times}
\usepackage{url}
\usepackage{graphics}
\usepackage{color}
\usepackage[pdftex]{hyperref}
\hypersetup{%
pdftitle={Your Title},
pdfauthor={Your Authors},
pdfkeywords={your keywords},
bookmarksnumbered,
pdfstartview={FitH},
colorlinks,
citecolor=black,
filecolor=black,
linkcolor=black,
urlcolor=black,
breaklinks=true,
}
\newcommand{\comment}[1]{}
\definecolor{Orange}{rgb}{1,0.5,0}

\begin{document}

\setlength{\paperheight}{11in}
\setlength{\paperwidth}{8.5in}
\setlength{\pdfpageheight}{\paperheight}
\setlength{\pdfpagewidth}{\paperwidth}

\title{An Interactive Machine Learning Framework}
\numberofauthors{3}
\author{
  \alignauthor Teng Lee\\
    \affaddr{University at Buffalo}\\
    \email{}
  \alignauthor James Johnson\\
    \affaddr{University at Buffalo}\\
    \email{}
  \alignauthor Steve Cheng\\
    \affaddr{University of California, Los Angeles}\\
    \email{}
}

\maketitle

\begin{abstract}
        Machine learning (ML) is believed to be an effective and efficient tool to build reliable prediction model or extract useful structure from an avalanche of data. However, ML is also criticized by its difficulty in interpretation and complicated parameter tuning. In contrast, visualization is able to well organize and visually encode the entangled information in data and guild audiences to simpler perceptual inferences and analytic thinking. But large scale and high dimensional data will usually lead to the failure of many visualization methods. In this paper, we close a loop between ML and visualization via interaction between ML algorithm and users, so machine intelligence and human intelligence can cooperate and improve each other in a mutually rewarding way. In particular, we propose "transparent boosting tree (TBT)", which visualizes both the model structure and prediction statistics of each step in the learning process of gradient boosting tree to user, and involves user's feedback operations to trees into the learning process. In TBT, ML is in charge of updating weights in learning model and filtering information shown to user from the big data, while visualization is in charge of providing a visual understanding of ML model to facilitate user exploration. It combines the advantages of both ML in big data statistics and human in decision making based on domain knowledge. We develop a user friendly interface for this novel learning method, and apply it to two datasets collected from real applications. Our study shows that making ML transparent by using interactive visualization can significantly improve the exploration of ML algorithms, give rise to novel insights of ML models, and integrates both machine and human intelligence.  
\end{abstract}



\section{Introduction}

Machine learning (ML) \cite{ML} plays an increasingly important role in data science today with great promises to improving our life. By fitting data to a flexible discriminative or generative model leveraging the relationship between variables and latent structures, ML is able to learn a prediction model or structured information that is consistent to the given data with statistical importance. The obtained prediction model or data structure can be applied to analysis and classification of new data instances. In this era, the avalanche of information gives rise to a growing interests in big data. While data collection becomes cheap, various current ML algorithms are designed for processing these large-scale and high dimensional data with noise with promising efficiency and accuracy. 

However, non-experts often treat ML as a fancy ``black box'' with few convincing interpretation of the model structure and learning process. In addition, even for ML researchers, how does the ML mechanism well-defined on statistical assumptions work on practical data, and how is the ML model updated in each iteration according to input features, are not transparent and usually ignored. The understanding and further exploration of ML models are highly limited in this case. Especially when applied to specific applications, it is essential to study when and why the ML algorithms work better or worse than expected, and then adjust the model accordingly. Therefore, a transparent ML tool that can clearly show the learning process to users can significantly facilitate the exploration of ML models.

\begin{figure*}[t!]
\begin{center}
 \includegraphics[width=1\linewidth]{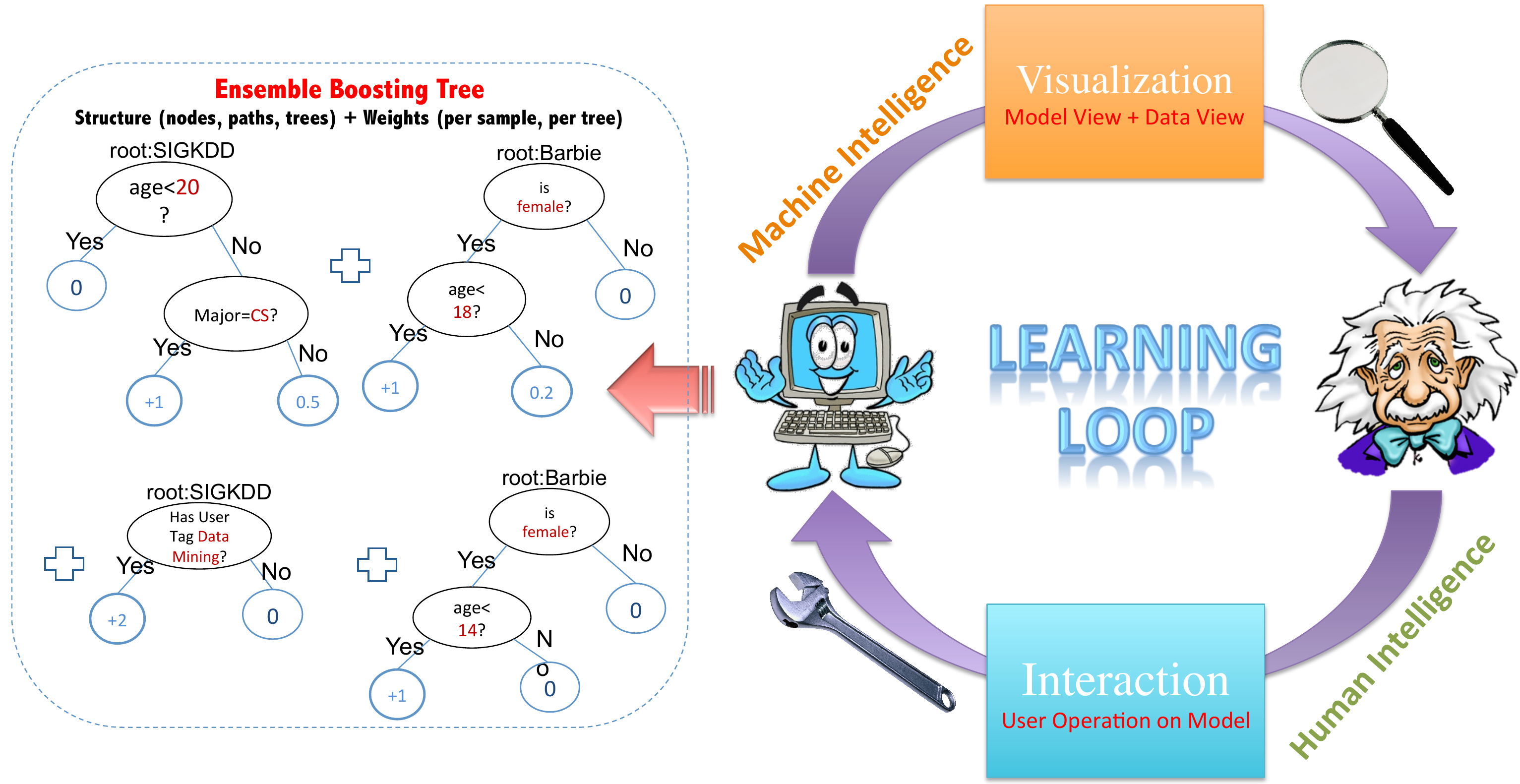}
\end{center}
   \caption{Interactive learning loop between machine learning and visualization by integrating both machine intelligence and human intelligence.}
\label{fig:system}
\end{figure*} 

Moreover, parameter tuning and determination of model complexity are two tricky problems that are critical to the performance of ML algorithms. Although there exists some statistical methods for solving them, it turns out that human intelligence based on domain knowledge always performs better and is more reliable on these decision-making problems, e.g., determining the depth of a decision tree, feature selection, and reasoning of performance enhancement. In addition, the emergence of human intelligence often offer better interpretation than machine intelligence. However, in the previous ML systems, it is also hard to involve users’ wisdom into learning process to improve both the generalization ability and interpretability of the ML models. 

In contrast, visualization \cite{visual} is a popular and powerful tool to explore data by concentrating human intelligence on visually encoded informations extracted from data. It shows different types of features of data in separate and interpretable ways, and also takes the relationship between features and samples into account. The visual representation of data is expected to help, facilitate and improve the comprehension of audiences, and thus stimulate more effective analytic thinking and decision making. However, visualization usually suffers from high dimensions and large volume of data. In this situation, the audiences suffer from an information overload caused by the big data, and the cognitive calculation could be slowed.

\subsection{Main Idea}

In this paper, we are going to close a loop between machine learning and visualization so they can benefit each other in data analysis and complete an interactive learning system that makes ML transparent to users. An overview of this system is given in Figure \ref{fig:system}. In each iteration of the loop, ML algorithm updates the prediction model according to given data and user feedback, then visualization provides a visual encoding to the model structures and the prediction results on training data by using the current model. In the visualization interface, user is able to track both the model structure and the prediction of training data, and operate on the model
to change its structures based on domain knowledge. These user operations will be the input of the next run of the ML algorithm. ML algorithm will update the model by treating user's feedback as constraint in its learning. In this loop, ML is in charge of updating model, choosing useful information to visualize, and processing user's feedback, while user is in charge of modification to the updated model. 

In particular, we choose gradient boosting tree (GBT) \cite{GBRT} as our ML algorithm in the loop because of its appealing performance in lots of practical problems. GBT is an ensemble of decision trees that are established in a forward stagewise fashion similar to boosting. It learns an additive model defining  decision rules for effective prediction from input features, such as classification and regression. GBT has been well known as the state-of-the-art method for many challenging real world problems. Each step of GBT learning can be treated as learning a decision tree based on all the previous trees. Each node of each decision tree corresponds to a selected feature and a decision rule. 

In the interactive learning system called ``transparent boosting tree (TBT)'', visualization interface provides four views of model and data, i.e., feature view listing the feature groups and highlighting the selected features, forest view listing all previous trees, tree view showing the detailed structure of the current tree, path purity view capturing the prediction results of training set on each node along a selected path on the tree, and history view tracking both the training error and test error of each update in the learning process. In each step, users are accessible to these views to evaluate the current model, and our system allows the users to add/remove a tree, select feature group for building a new tree, remove a node on the current tree, expand a leaf node, and go back to any previous model in the learning loop. Then GBT algorithm will take these user operations as inputs to update the ensemble of decision trees accordingly. 

In this paper, we implement TBT by building an interactive visualization system, and apply it to two practical datasets, mushroom and fusion health. The former aims at justifying if a given mushroom is poisonous or not, while the latter aims at predicting the healthy status of a patient. TBT is demonstrated to be a promising tool for exploration of GBT, visualization of ML model, and human-ML interaction.

\section{Related Works}

There are several, though not many, early efforts on improving the interpretation of ML models by using effective visualization \cite{aai}. The visualization is critical in many systems, e.g. health \cite{huang2014ezwakeup,liu2015breathsens}, medical \cite{huang2014using}, and data mining \cite{witten2005data}.

EnsembleMatrix \cite{Ematrix} enables users to choose any combination of multiple classifiers to build a combination model, after showing users the confusion matrix of all classifiers. Thus it provides an interaction visualization tool emerging user intelligence in ML process. However, both the training and mode structure of each classifier are still invisible to users.

Some recent works like BigML\footnote{https://bigml.com} focuses on the visualization of decision tree, which is related to ensemble boosting tree in our project. However, decision tree is less powerful and also simpler than ensemble boosting tree. In addition, they compute all possible paths before showing them to users rather than instantaneously update them, and allow fewer operations of users on the tree. So the loop has not been really built, the benefits of interaction is limited, and the computation could be expensive.

\section{Transparent Boosting Tree}

\begin{figure*}[t!]
\begin{center}
 \includegraphics[width=1.05\linewidth]{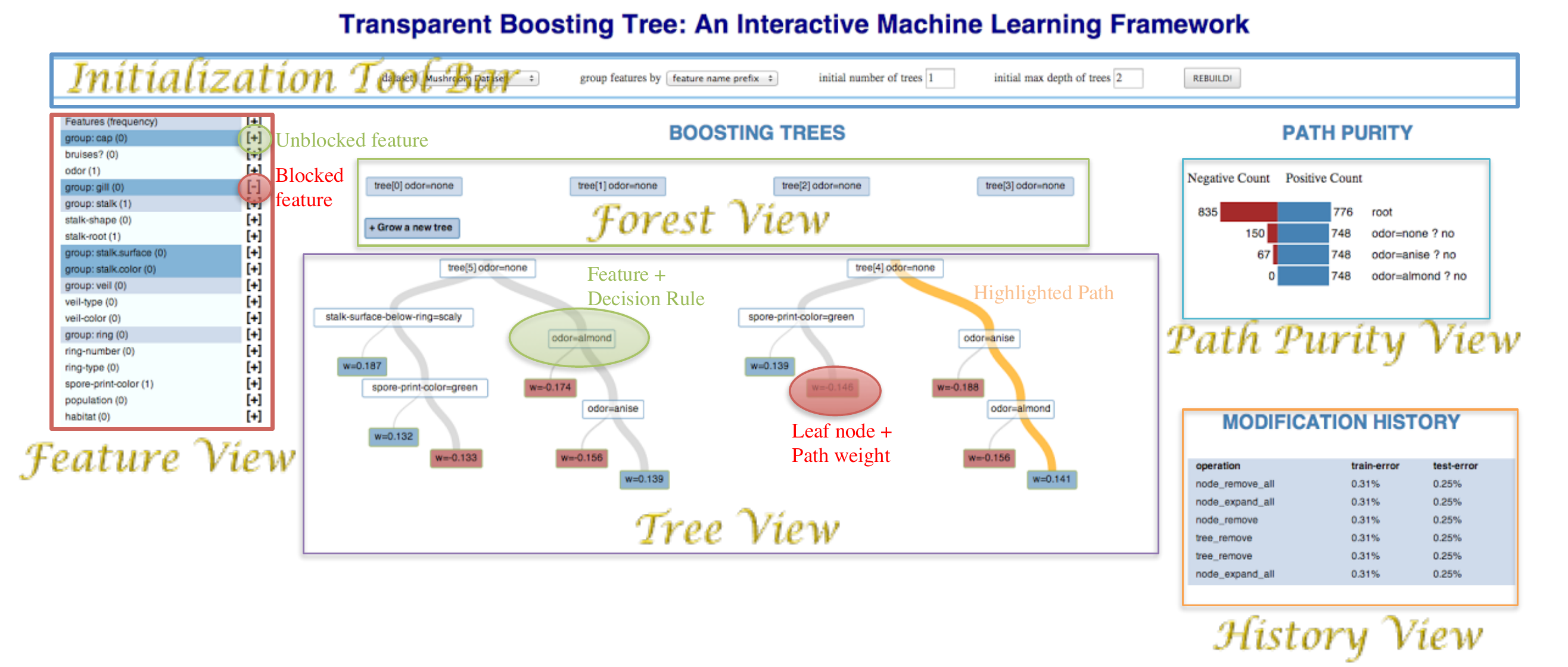}
\end{center}
   \caption{Visual encoding and exploration in five views of visualization interface of transparent boosting tree.}
\label{fig:interface}
\end{figure*} 

\subsection{Learning Algorithm for Gradient Boosting Tree}

One goal of TBT is to visualize the learning process and the model structure of GBT. GBT builds additive model $F(x)$ composed of $M$ decision trees $h_m(x)$ with input features $x$, i.e., 
\begin{equation}
F(x)=\sum_{m=1}^M \gamma_mh_m(x).
\end{equation}
Each decision tree $h_m(x)$ is a weak learner that can handle data of mixed type and to model complex nonlinear function. Similar to boosting, GBT builds the additive model in a forward stagewise fashion such that
\begin{equation}
F_m(x)=F_{m-1}(x)+\gamma_mh_m(x).
\end{equation}
In each stage, fixing the previous ensemble of trees $F_{m-1}(x)$, the current decision tree $h_m(x)$ is chosen to minimize the loss function $L$.
\begin{equation}
h_m(x)=\arg\min_h\sum_{i=1}^n L(y_i, F_{m-1}(x_i)-h(x_i)).
\end{equation}
This minimization problem can be solved by using a steepest gradient descent, which is
\begin{equation}
F_m(x)=F_{m-1}(x)+\gamma_m\sum_{i=1}^n\nabla L(y_i, F_{m-1}(x_i)),
\end{equation}
where the step size $\gamma_m$ is determined by line search. The final additive model is usually represented by the combination of all paths on all the trees with different weight per path.

In learning of each decision tree, the gradient $\nabla L(y_i, F_{m-1}(x_i))$ is firstly assigned to each training sample as its importance weight in learning the current tree. Then the nodes and their corresponding decision rules defining the tree are selected in a greedy manner from root to leaves. Specifically, for each node, learning algorithm chooses the best threshold for each available feature by evaluating all possible thresholds. The selected threshold defines the best decision rule associated to each feature. By using both the importance weights and Hessian matrix, we can thus compute the information gain caused by applying each feature and its decision rule on the node. The pair which results in the maximal information gain is selected to define the current node. 

In TBT, we involve the user operations into the learning process of GBT. In particular, TBT allows users to operate on the features and trees that used to be selected by using greedy method, and to determine the number of trees and the depth of a path. Thus the user feedback can be treated as a human regularization or constraint to limit the model complexity of GBT, and to correct the possible mistakes made by pure greedy selection. This is expected to be effective in avoiding overfitting, simplifying model, and improving intepretability. Because human with domain knowledge is believed to be good at those decision making tasks. 

We let the GBT learning algorithm in charge of the update of thresholds and importance weights, because computer is good at searching and optimizing these numerical variables. Therefore, the advantages of both human intelligence and machine intelligence can be combined in TBT.

\subsection{Design of Interactive Visualization Interface}

The visualization interface of TBT is composed of one initialization tool bar and five views, i.e., feature view listing the feature groups and highlighting the enabled features, forest view listing all previous trees, tree view showing the detailed structure of current tree, path purity view capturing the prediction results of training set on each node along a selected path on the tree, and history view tracking both the training error and test error of each update in the learning process. 

In the initialization tool bar, user is able to choose the dataset to analyze, the method to group features in the feature view, the initial number of trees in GBT, the initial maximal depth of each tree. Clicking the ``REBUILD!'' button will provide an initial GBT model, which is learned by original GBT algorithm and visualized in the five views.

In the following tables, we list the major visually encoded information and interaction/user operation in each view.

\begin{table}[ht]
\caption{Visual encoding and user operation in feature view.}
\begin{center}
\begin{tabular}{|l|}
\hline
Visual Encoding/Exploration\\
\hline
Feature group (text)\\
Feature name (text)\\
No. of features per group (integer) \\
\hline
\hline 
User Operation/Interaction\\
\hline
allow the feature (right click feature \& select)\\
block the feature (right click feature \& select)\\
\hline
\end{tabular}
\end{center}
\end{table}

\begin{table}[ht]
\caption{Visual encoding and user operation in forest view.}
\begin{center}
\begin{tabular}{|l|}
\hline
Visual Encoding/Exploration\\
\hline
Tree index (integer)\\
Feature and decision rule for root node (text)\\
\hline
\hline 
User Operation/Interaction\\
\hline
Collapse the whole tree (left click)\\
\hline
\end{tabular}
\end{center}
\end{table}

\begin{table}[ht]\small
\caption{Visual encoding and user operation in tree view.}
\begin{center}
\begin{tabular}{|l|}
\hline
Visual Encoding/Exploration\\
\hline
Feature (text)\\
Decision rule (text)\\
Path weight on leaf node (ratio)\\
Major class on leaf node (color)\\
No. of samples on each edge (width)\\
Highlight a path (brushing a path)\\
Link a path to its prediction bar charts\\
\hline
\hline 
User Operation/Interaction\\
\hline
Remove the current tree (right click root node \& select)\\
Grow a new tree (right click root node \& select)\\
Remove the node on the current tree (right click node \& select)\\
Remove the node on all trees (right click node \& select)\\
Expand the leaf node on the current tree (right click node \& select)\\
Expand the leaf node on all trees (right click node \& select)\\
\hline
\end{tabular}
\end{center}
\end{table}
\normalsize

\begin{table}[ht]
\caption{Visual encoding in path purity view.}
\begin{center}
\begin{tabular}{|l|}
\hline
Visual Encoding/Exploration\\
\hline
Prediction statistics on a path (linked to brushed path)\\
Negative and positive counts on nodes (bar charts)\\
Decision rule (text)\\
\hline
\end{tabular}
\end{center}
\end{table}

\begin{table}[ht]
\caption{Visual encoding and user operation in history view.}
\begin{center}
\begin{tabular}{|l|}
\hline
Visual Encoding/Exploration\\
\hline
Historical operation (text)\\
Training error (ratio)\\
Test error (ratio)\\
\hline
\hline 
User Operation/Interaction\\
\hline
Go back to former model (right click row \& select restore)\\
\hline
\end{tabular}
\end{center}
\end{table}

On one hand, the ``forest view'' and ``tree view'' visualize the structure of GBT model in learning process, i.e., multiple decision trees from root node to current leaf nodes, the weight of each path, and the feature with associated decision rule on each node. These two views provide a detailed and transparent visualization of GBT model structure to user. They lead to a clear explanation of the current model, e.g., what features were selected to build decision rules, and how these features cooperate with each other. So users can justify which nodes on the current model are consistent with domain knowledge or common sense, and which are not. 

On the other hand, ``path purity view'' and ``history view'' visualize the prediction results on each node along the brushed path in the tree view, and the overall historical training/test error after each user operation, respectively. These two views let users track the performance changes caused by the feature(s) and the associated decision rule(s) on each node, each pat, each tree, and each user operation in the history. Even without sufficient domain knowledge, users can still find important clues from the prediction results of the current GBT model to evaluate the performance of selected features, decision rules, established trees and paths. 

By showing users both the model structure and performance on training data, we expect to collect the wisdom of users from both their domain knowledge and judgement based on the prediction result on data via their feedback. In our visualization tool, users have complete freedom to modify the structure of each tree, e.g., to choose feature for each node, to prune an existing node/path, and to add a new node/path. The model will be updated according to these instantaneous feedbacks from users, and we will show the new views for the updated model. 

In the learning process, ML algorithm is in charge of learning candidates of model, estimating the numerical weights of the model, and updating model according to user feedback, while users are in charge of modifying the ML model by their domain knowledge and feedback from the current learning results. Interactive visualization bridges these two components by conveying the feedback of one to the other. Therefore,  we can close a loop between these two so that one helps enhance the performance of the other in a mutually rewarding way, until we achieve a promising ML model.

\subsection{Exploration of GBT model in TBT}

\begin{figure*}[t!]
\begin{center}
 \includegraphics[width=1.05\linewidth]{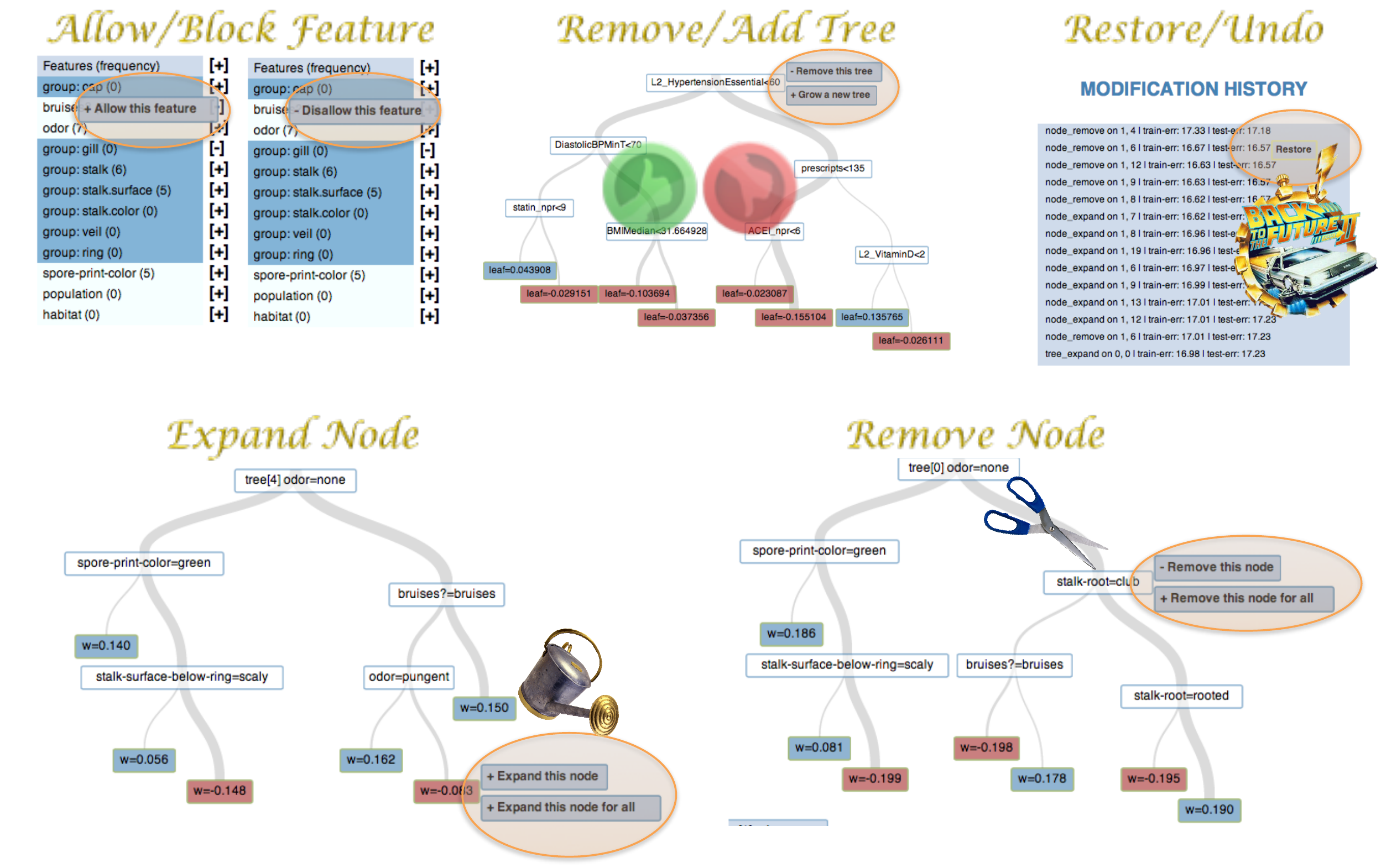}
\end{center}
   \caption{Interaction and user operation in visualization interface of transparent boosting tree.}
\label{fig:user}
\end{figure*} 

One primary contribution of this paper is that TBT enables full exploration of complex machine learning model by user. We show the visual encoding/exploration contained in TBT in Figure \ref{fig:interface}. In particular, although TBT allows to learn hundreds of trees in a second, the detailed structure of each decision tree can be collapsed by an animation and shown in the tree view when clicking the tree index listed in the forest view. Thus user is able to evaluate each tree by going to its details quickly. 

In the tree view, user is able to explore the feature and corresponding decision rule on each node, as well as the composite decision rule defined by each path, and the number of training samples associated with each edge of each path. So the features and decision rules can be evaluated by users' domain knowledge. In addition, by brushing arbitrary path on the tree, the prediction statistics on each node along the selected path can be displayed in bar charts within the path purity view. By investigating the prediction statistics, it is very convenient for user to determine the best depth of the path that can avoid possible overfitting, and do manual feature selection according to the contribution of each feature in the prediction task.

Therefore, users is able to gain a comprehensive and course-to-fine understanding of the ML model via the visualization and exploration of both the model structure and prediction results on data. This is essentially helpful to generating feedback of high quality after analytic thinking and cognitive inference in human brain. 

\subsection{Interaction and User Operation in TBT}

Another key insight of this paper is that TBT allows rich user operations on the ML model and is able to take those user feedback as input of the learning algorithm in the next model update. As shown in Figure \ref{fig:user}, users are allowed to operate on different scales of GBT model. 

Specifically, on the forest scale, they can remove/add arbitrary number of trees to the model. In addition, in any stage of learning process, user can define a subset of features by ``allow/block features'' in the feature view, such that all any feature selected in the rest trees are confined in the subset. 

On the tree scale, users can remove any node or expand any leaf node. This results in feature selection. It worth noting that TBT supports a very powerful and useful function called ``remove/expand this node for all'', which is able to remove/add the selected node on all the same paths appearing in all the trees. This plays a role of global filter that reflects the attitude of users to the composite decision rule defined by the path.

On the time scale, the ``restore'' button in history view of TBT allows user to go back to the model in any learning stage marked by a historical user operation. This is very necessary because some human operations are possible to result in a weaker model. By tracking the training/test error in the history view, users are able to justify if the operations are helpful to the learning process or not. And the ``restore'' button can conveniently undo the operations that are justified not helpful, while users do not need to go back the the very beginning of the learning process.

\section{Results}

We apply TBT to two datasets, i.e., mushroom\footnote{https://archive.ics.uci.edu/ml/datasets/Mushroom} and fusion health Diabetes dataset\footnote{http://www.kaggle.com/c/pf2012-diabetes/data}. In mushroom dataset, there are $8124$ data samples with $22$ features, and the goal is to train a classifier separating poisonous mushrooms and edible mushrooms. In fusion health dataset, there are $9948$ training samples and $4979$ test samples, and the goal is to recognize if a given patient has a diagnosis of Type 2 diabetes mellitus (T2DM). TBT is going to build GBT model for prediction tasks for the two datasets. 

It shows that TBT performs effectively in presenting a transparent and interpretable understanding of GBT model and its learning process to users. In spite of the large number of samples and features, TBT is able to show how the GBT model is established by integrating different features and decision rules using ensemble of trees, and how the model works on real data. User is able to explore the model from forest scale to feature scale. Moreover, the interaction and user operation is proved to be effective in testing the performance and tracking the changes casued of each part of the model. Furthermore, the update of model in TBT is real-time scale, and thus the learning loop between ML and visualization can be iterated efficiently.

\section{Discussion}

By applying TBT to the two real datasets, we discover that it can provide several novel insights that cannot be offered by directly running GBT on the datasets. Firstly, instead of learning a great number of decision trees blindly, we find out that lots of decision trees learned in GBT share the same paths or even have very similar structure. This indicates that the complexity of GBT can be significantly reduced by user exploration.

Secondly, TBT is able to show the features or feature groups that have been selected to build decision rules in the trees. The prediction statistics along a path in path purity view and training/test error shown in history view provide important clues to evaluate and understand the roles the selected features and decision rules play in the prediction. 

Thirdly, by using the information shown in the path purity view and history view, it is convenient to prune the trees and thus avoid overfitting of GBT model.

\section{Future Work}

In the future, we will do user study on our published website and collect the performance of TBT by involving different users' interaction into the learning process. The collected data can be used to evaluate TBT more accurately. 

Moreover, we plan to study whether and when human intelligence in TBT is able to improve the performance of ML in the loop. Both a theoretical analysis and empirical studies are necessary to be conducted in the future.

Furthermore, we will attempt to extend the transparent machine learning idea to other ML models and algorithms, and try to find if the interactive visualization can result in new understanding of the popular ML black boxes.

\bibliographystyle{abbrv}
\bibliography{sample}

\end{document}